\documentclass[sigconf]{acmart}

\AtBeginDocument{%
  \providecommand\BibTeX{{%
    \normalfont B\kern-0.5em{\scshape i\kern-0.25em b}\kern-0.8em\TeX}}}

\copyrightyear{2022}
\acmYear{2022}
\setcopyright{acmcopyright}\acmConference[CIKM '22]{Proceedings of the 31st ACM
International Conference on Information and Knowledge Management}{October
17--21, 2022}{Atlanta, GA, USA}
\acmBooktitle{Proceedings of the 31st ACM International Conference on Information
and Knowledge Management (CIKM '22), October 17--21, 2022, Atlanta, GA, USA}
\acmPrice{15.00}
\acmDOI{10.1145/3511808.3557266}
\acmISBN{978-1-4503-9236-5/22/10}

\usepackage{multicol}
\usepackage{multirow}
\usepackage{graphics}
\usepackage{xcolor}
\usepackage{algorithm}
\usepackage{algorithmicx}
\usepackage{overpic}
\usepackage{algpseudocode}
\usepackage{subcaption}
\usepackage{bbding}
\usepackage{enumitem}
\usepackage{balance}
\usepackage{hyperref}
\begin{document}
%%\fancyhead{}

\title{Contrastive Learning with Bidirectional Transformers for Sequential Recommendation}


%%\author{Hanwen Du$^{1}$, Hui Shi$^{1}$, Pengpeng Zhao$^{1^*}$, Deqing Wang$^{2}$,\\ Victor S.Sheng$^{3}$, Yanchi Liu$^{4}$, Guanfeng Liu$^{5}$ and Lei Zhao$^{1}$} \thanks{$^*$Corresponding author.}

%%\affiliation{%
%%  \institution{$^{1}$Soochow University, $^{2}$Beihang University, $^{3}$Texas Tech University, $^{4}$Rutgers University, $^{5}$Macquarie University}
%%  \country{}
%}
%\email{{hwdu, hshi1}@stu.suda.edu.cn, {ppzhao, zhaol}@suda.edu.cn, dqwang@buaa.edu.cn} \email{victor.sheng@ttu.edu, yanchi.liu@rutgers.edu, guanfeng.liu@mq.edu.au}
\author{Hanwen Du}
\email{hwdu@stu.suda.edu.cn}
\affiliation{%
\institution{Soochow University}
\city{Suzhou}
\state{Jiangsu}
\country{China}
}

\author{Hui Shi}
\email{hshi1@stu.suda.edu.cn}
\affiliation{%
\institution{Soochow University}
\city{Suzhou}
\state{Jiangsu}
\country{China}
}

\author{Pengpeng Zhao$^{^*}$}
\thanks{$^*$Corresponding author.}
\email{ppzhao@suda.edu.cn}
\affiliation{%
\institution{Soochow University}
\city{Suzhou}
\state{Jiangsu}
\country{China}
}

\author{Deqing Wang}
\email{dqwang@buaa.edu.cn}
\affiliation{%
\institution{Beihang University}
\city{Beijing}
\country{China}
}

\author{Victor S. Sheng}
\email{victor.sheng@ttu.edu}
\affiliation{%
\institution{Texas Tech University}
  \city{Lubbock}
  \state{Texas}
  \country{USA}
}

\author{Yanchi Liu}
\email{yanchi.liu@rutgers.edu}
\affiliation{%
\institution{Rutgers University}
\city{New Brunswick}
\state{New Jersey}
\country{USA}
}

\author{Guanfeng Liu}
\email{guanfeng.liu@mq.edu.au}
\affiliation{%
\institution{Macquarie University}
\city{Sydney}
\country{Australia}
}

\author{Lei Zhao}
\email{zhaol@suda.edu.cn}
\affiliation{%
\institution{Soochow University}
  \city{Suzhou}
  \state{Jiangsu}
  \country{China}
}

\renewcommand{\shortauthors}{Hanwen Du et al.}

\begin{abstract}
Contrastive learning with Transformer-based sequence encoder has gained predominance for sequential recommendation. It maximizes the agreements between paired sequence augmentations that share similar semantics. However, existing contrastive learning approaches in sequential recommendation mainly center upon left-to-right unidirectional Transformers as base encoders, which are suboptimal for sequential recommendation because user behaviors may not be a rigid left-to-right sequence. To tackle that, we propose a novel framework named \textbf{C}ontrastive learning with \textbf{Bi}directional \textbf{T}ransformers for sequential recommendation (\textbf{CBiT}). Specifically, we first apply the slide window technique for long user sequences in bidirectional Transformers, which allows for a more fine-grained division of user sequences. Then we combine the cloze task mask and the dropout mask to generate high-quality positive samples and perform multi-pair contrastive learning, which demonstrates better performance and adaptability compared with the normal one-pair contrastive learning. Moreover, we introduce a novel dynamic loss reweighting strategy to balance between the cloze task loss and the contrastive loss. Experiment results on three public benchmark datasets show that our model outperforms state-of-the-art models for sequential recommendation. Our code is available at this link: \url{https://github.com/hw-du/CBiT/tree/master}.
\end{abstract}





\maketitle

\section{Introduction}
Sequential recommendation system aims to model the dynamic preferences in users' historical interactions and predicts the subsequent items that users will probably interact with in the future. Traditional methods \cite{MDP,FPMC,Fossil} are based on the Markov Chain (MC) assumption that the next item only depends on previous items. With the advancements in deep learning in the recent past, various models employ deep neural networks, such as Convolutional Neural Networks (CNNs) \cite{tang2018caser} and Recurrent Neural Networks (RNNs) \cite{srnn2016,Hidasi_2018}, as base sequence encoders to generate hidden representations of sequences. The limitations are that CNNs are only effective in capturing local features \cite{tang2018caser} while RNNs display poor parallelism capacity \cite{rnnanalysis}. Recently, Transformers \cite{vaswani2017transformer} have emerged as a powerful architecture in various research fields. Different from CNNs or RNNs, the self-attention mechanism in Transformers can automatically assign attention weights to items at different positions, which is capable of capturing both global and local features and can also be effectively trained through parallel computation. Thus, various sequential models \cite{kang18attentive,LSSA,Sun2019bert,Xu2020Contrastive,liu2021contrastive,DuoRec} adopt Transformers as sequence encoder to capture item correlations via the self-attention mechanism and obtain high-quality sequence representations.

Despite effectiveness, current Transformer-based models suffer from data sparsity and interaction noise, owing to the extreme sparsity of interactions. To alleviate these issues, contrastive learning has been introduced to Transformer-based models for sequential recommendation. As a paradigm of self-supervised learning, contrastive learning demonstrates strong ability in generating high-quality embedding representations from unlabelled data, via maximizing the agreements between positive samples and pushing negative samples apart from positive samples. For example, CL4SRec \cite{Xu2020Contrastive} and CoSeRec \cite{liu2021contrastive} construct positive samples from an original sequence through data augmentations to perform contrastive learning. Duorec \cite{DuoRec} constructs positive samples via unsupervised dropout and supervised positive sampling. Such contrastive paradigm can enhance the discriminating ability of sequence encoders as well as improve robustness and noise resistance.

However, all existing works \cite{Xu2020Contrastive,liu2021contrastive,DuoRec} devise contrastive sequential recommendations based on unidirectional Transformers, neglecting the superiority of bidirectional Transformers. The limitation is that unidirectional Transformers can only consider information from left to right. Yet in the real world, user behaviors may not be a rigid left-to-right sequence. Various external factors \cite{youtuberec} disorganize the original sequence order, and skip behaviors \cite{tang2018caser} also exist in sequential patterns. By comparison, the attention mechanism in the bidirectional Transformers models items from both sides and incorporates contextual information from both directions, resulting in better performance \cite{Sun2019bert} than its unidirectional counterpart such as SASRec \cite{kang18attentive}. We argue that contrastive learning based on such left-to-right unidirectional Transformers are suboptimal for sequential recommendation because user behaviors may not be a rigid left-to-right sequence, so it is meaningful to devise contrastive learning frameworks based on the architecture of bidirectional Transformers.

In fact, it is not trivial to introduce the contrastive paradigm into the architecture of bidirectional Transformers. The characteristics of bidirectional Transformers and contrastive learning need to be carefully examined to achieve a balance, which requires us to answer the following  questions: (1) \emph{How to choose an augmentation strategy for contrastive learning?} A simple answer is to explicitly apply data augmentation similar to \cite{Xu2020Contrastive,liu2021contrastive}, such as masking items. Yet such data augmentation strategy collides with the training objective of bidirectional Transformers---the cloze task, which also masks items.
(2) \emph{How to construct reasonable positive samples for contrastive learning?} It is hard to find a reasonable self-supervision signal that could indicate whether selected samples are semantically similar or dissimilar. Even if a pair of data-augmented samples stem from the same original sequence, considering them as a pair of positive samples may still be unreasonable because data augmentation may corrupt the original semantics.

To tackle the challenges mentioned above, we devise contrastive learning approaches suitable for the properties of bidirectional Transformers and 
propose a novel framework named \textbf{C}ontrastive learning with \textbf{Bi}directional \textbf{T}ransformers for sequential recommendation (\textbf{CBiT}). Specifically, we first apply the slide window technique for long sequences in bidirectional Transformers to resolve the restriction of maximum sequence length, which preserves all the training data and helps bidirectional Transformers capture more fine-grained features. Then we combine both the cloze task mask and the dropout mask to generate a collection of positive samples and extrapolate the normal one-pair contrastive learning to multi-pair instances. Compared with one-pair contrastive learning, multi-pair contrastive learning not only provides harder samples which benefits high-order feature extraction, but also alleviates the negative impact of false negative samples by incorporating more positive samples. Moreover, we design a dynamic loss reweighting strategy, which dynamically calculates the proportion of the contrastive loss, to speed up convergence and further improve performance. Extensive experiments on three public benchmark datasets confirm the effectiveness of our framework. The contributions of our paper can be summarized as follows:

\begin{itemize}
[leftmargin =  8pt,topsep=1pt]
    \item We propose a novel framework of contrastive learning based on bidirectional Transformers in sequential recommendation. To the best of our knowledge, we are the first to introduce contrastive learning under the architecture of bidirectional Transformers in sequential recommendation.
    \item We use both the cloze task mask and the dropout mask as augmentations to generate positive samples and extrapolate the normal one-pair contrastive learning to multi-pair instances. We also propose a novel dynamic loss reweighting strategy to smooth multi-pair contrastive loss.
    \item We conduct extensive experiments on three public benchmark datasets to verify the effectiveness of our approach.
\end{itemize}
\section{RELATED WORK}
\subsection{Sequential Recommendation}
Early works on sequential recommendation employ Markov Chains (MCs) to capture the dynamic transition of user interactions, such as MDP \cite{MDP}, FPMC \cite{FPMC} and Fossil \cite{Fossil}. Later, RNNs such as Gated Recurrent Unit (GRU) \cite{gru} are introduced into sequential recommendation to model user interactions \cite{srnn2016,Hidasi_2018,HRNN}. CNNs are also proved effective in modeling short-term user interests \cite{tang2018caser}. Besides, other algorithms such as reinforcement learning \cite{RouteOptimization,intelligent,sasrecsqnsac} have also shown promising results in sequential recommendation.

Recently, the success of Transformers \cite{vaswani2017transformer} in natural language processing \cite{Devlin2019BERT} and computer vision \cite{dosovitskiy2020vit} brings focus on the possibility of using Transformers as sequence encoders in sequential recommendation. SASRec \cite{kang18attentive} employs unidirectional Transformers fulfill the next-item prediction task in sequential recommendation. BERT4Rec \cite{Sun2019bert} improves SASRec by using bidirectional Transformers and a cloze task in sequential recommendation. LSSA \cite{LSSA} proposes a long- and short-term self-attention network to consider both long-term preferences and sequential dynamics. SR-GNN \cite{SR-GNN} and GC-SAN \cite{GCSAN} combine Graph Neural Networks (GNNs) with self-attention networks to capture both local and global transitions. FDSA \cite{FDSA} and $\rm{S}^3$-Rec \cite{CIKM2020-S3Rec} employ Transformers to fuse context data into sequential recommendation. CL4SRec \cite{Xu2020Contrastive}, CoSeRec \cite{liu2021contrastive} and DuoRec \cite{DuoRec} add additional contrastive learning modules to Transformers to enhance the quality of sequence representations.
\subsection{Self-Supervised Learning}
Self-supervised learning aims to extract contextual features from unlabeled data \cite{Bengio2021DeepLF}. Advancements in various research fields, including natural language processing \cite{gao2021simcse,liang2021rdrop,SCLlanguage,UDA}, computer vision \cite{chen2020simple,mocov1,mocov2,mocov3} and recommender systems \cite{ExploitingAestheticPreference,Sun2019bert,CIKM2020-S3Rec,Xu2020Contrastive,liu2021contrastive,MMInfoRec,DuoRec} have demonstrated the potential of self-supervised methods in capturing complex features and obtaining high-quality representations.

Self-supervised learning can be classified into two broad categories, i.e., the generative paradigm and the contrastive paradigm \cite{generativeorcontrastive}. For example, the cloze task in BERT4Rec \cite{Sun2019bert} is a generative method where the model learns to predict part of its input. Many works have also confirmed the effectiveness of the contrastive paradigm in sequential recommendation \cite{CIKM2020-S3Rec,Xu2020Contrastive,liu2021contrastive,DuoRec}. $\rm{S}^3$-Rec \cite{CIKM2020-S3Rec} adopts a two-stage strategy to incorporate contextual information into sequential recommendation through self-supervised methods such as mutual information maximization. CL4SRec \cite{Xu2020Contrastive} uses three data augmentation methods suitable for sequential recommendation to generate positive samples for contrastive learning. CoSeRec \cite{liu2021contrastive} improves CL4SRec by introducing robust data augmentation methods. DuoRec \cite{DuoRec} performs contrastive learning from the model level by introducing supervised and unsupervised objectives for contrastive learning. Different from these works, our work devises contrastive learning methods based on the architecture of bidirectional Transformers.

\section{Preliminaries}
\subsection{Problem Statement}
In sequential recommendation, we denote $  \mathcal{U}=\{u_1,u_2,\cdots\,u_{\lvert \mathcal{U} \rvert}\}$ as a set of users, $  \mathcal{V}=\{v_1,v_2,\cdots\,v_{\lvert \mathcal{V} \rvert}\}$ as a set of items, and $s_u=\{v^{(u)}_1,v^{(u)}_2,\cdots\,v^{(u)}_{\lvert s_u \rvert}\}$ as an interaction sequence sorted in the chronological order, where $v^{(u)}_i\in \mathcal{V}$ denotes the item that user $u$ has interacted with at the $i$-th timestamp. The task of sequential recommendation is to predict the next item that user $u$ is probably interested in, and it can be formulated as generating the probability of all items for user $u$ at the next timestamp $\lvert s_u \rvert +1$:\begin{displaymath}p(v^{(u)}_{\lvert s_u\rvert+1}=v|s_u).\end{displaymath}
\subsection{Bidirectional Transformers}
The architecture of bidirectional Transformers incorporates a stack of Transformer blocks. Each Transformer block consists of a multi-head self-attention module and a feed-forward network. Multiple Transformer blocks are stacked together as a deep network.
\subsubsection{Multi-Head Self-Attention.} Multi-head self-attention is effective to extract information from $h$ different subspaces at different positions \cite{Devlin2019BERT,vaswani2017transformer}.  Given the hidden representation ${\emph{\textbf{H}}}^{\thinspace l}\in {\mathbb{R}}^{T\times d}$ for the $l$-th layer with maximum sequence length $T$ and hidden dimensionality $d$, the computation is formulated as follows:
\begin{equation}
\begin{aligned}
\rm{MH}({\emph{{\textbf{H}}}}^{\emph{\thinspace l}}\thinspace)&=\rm{concat({head}_1;{head}_2;\cdots;{head}_{\emph{h}})}{\emph{\textbf{W}}^{\thinspace O}},\\
{\rm{head}}_i&=\rm{Attention}({\emph{{\textbf{H}}}}^\emph{\thinspace l}{\emph{{\textbf{W}}}}_{i}^{\emph{\thinspace Q}},{\emph{{\textbf{H}}}}^\emph{\thinspace l}{\emph{{\textbf{W}}}}_{i}^{\emph{\thinspace K}},{\emph{{\textbf{H}}}}^\emph{\thinspace l}{\emph{{\textbf{W}}}}_{i}^{\emph{\thinspace V}}),
\end{aligned}
\end{equation}
where ${\emph{{\textbf{W}}}}_{i}^{\emph{\thinspace Q}}\in{\mathbb{R}}^{d\times d / h}$, ${\emph{{\textbf{W}}}}_{i}^{\emph{\thinspace K}}\in{\mathbb{R}}^{d\times d / h}$, ${\emph{{\textbf{W}}}}_{i}^{\emph{\thinspace V}}\in{\mathbb{R}}^{d\times d / h}$ and ${\emph{{\textbf{W}}}}^{\emph{\thinspace O}}\in{\mathbb{R}}^{d\times d}$ are learnable parameters. The attention mechanism is implemented by scaled dot-product and softmax operation:
\begin{equation}
\rm{Attention}(\emph{\textbf{Q}},\emph{\textbf{K}},\emph{\textbf{V}}\thinspace)=\rm{softmax}(\frac{\emph{\textbf{Q}}\emph{\textbf{K}}^{\top}}{\sqrt{\emph{d}/\emph{h}}})\emph{\textbf{V}},
\end{equation}
where $\emph{\textbf{Q}},\emph{\textbf{K}},\emph{\textbf{V}}$ stand for query, key, value respectively and $\sqrt{d/h}$ is a scale factor to avoid large values of the inner product.
\subsubsection{Feed-Forward Network.}
Given that multi-head self-attention is mainly based on linear projections, adding a feed-forward network after the attention layer is conducive to capturing non-linear features. It can be formulated as follows:
\begin{equation}
\begin{aligned}
\mathrm{PFFN}({\emph{{\textbf{H}}}}^{\emph{\thinspace l}})&=[\mathrm{FFN}({\emph{{\textbf{h}}}}_{1}^{\emph{l}})^{\top};\mathrm{FFN}({\emph{{\textbf{h}}}}_{2}^{\emph{l}})^{\top};\cdots;\mathrm{FFN}({\emph{{\textbf{h}}}}_{\lvert s_{u} \rvert}^{\emph{l}})^{\top}],\\
\mathrm{FFN}({\emph{{\textbf{h}}}}_{i}^{\emph{l}})&=\mathrm{GeLU}({\emph{{\textbf{h}}}}_{i}^{\emph{l}}{\emph{{\textbf{W}}}}_{1}+{\emph{{\textbf{b}}}}_{1}){\emph{{\textbf{W}}}}_{2}+{\emph{\textbf{b}}}_{2},
\end{aligned}
\end{equation}
where ${\emph{{\textbf{W}}}}_{1} \in {\mathbb{R}}^{d \times 4d}$,${\emph{{\textbf{W}}}}_{2} \in {\mathbb{R}}^{4d \times d}$,${\emph{{\textbf{b}}}}_{1} \in {\mathbb{R}}^{4d} $and $ {\emph{{\textbf{b}}}}_{2} \in {\mathbb{R}}^{d}$ are trainable parameters shared across all positions.
\subsubsection{Interlayer Stacking.} Multiple Transformer blocks are stacked together to construct a deep network. Mechanisms including residual connection \cite{he2016deep}, layer normalization \cite{ba2016layer} and dropout \cite{dropout} are introduced between layers to prevent overfitting. In summary, the bidirectional Transformer encoder $\mathtt{Trm}$ can be defined as follows:
\begin{equation}
\begin{aligned}
    \mathtt{Trm}({\emph{{\textbf{H}}}}^{\thinspace l})&=\mathrm{LayerNorm}({\emph{{\textbf{F}}}}^{\thinspace l}+\mathrm{Dropout}(\mathrm{PFFN}({\emph{{\textbf{F}}}}^{\thinspace l}))),\\
    {\emph{{\textbf{F}}}}^{\thinspace l}&=\mathrm{LayerNorm}({\emph{{\textbf{H}}}}^{\thinspace l}+\mathrm{Dropout}(\mathrm{MH}({\emph{{\textbf{H}}}}^{\thinspace l}))).
\end{aligned}
\end{equation}

\section{Proposed Framework}
\begin{figure}
 \centering
 \begin{overpic}[width=0.92\textwidth,trim = 0 112 235 0,clip]{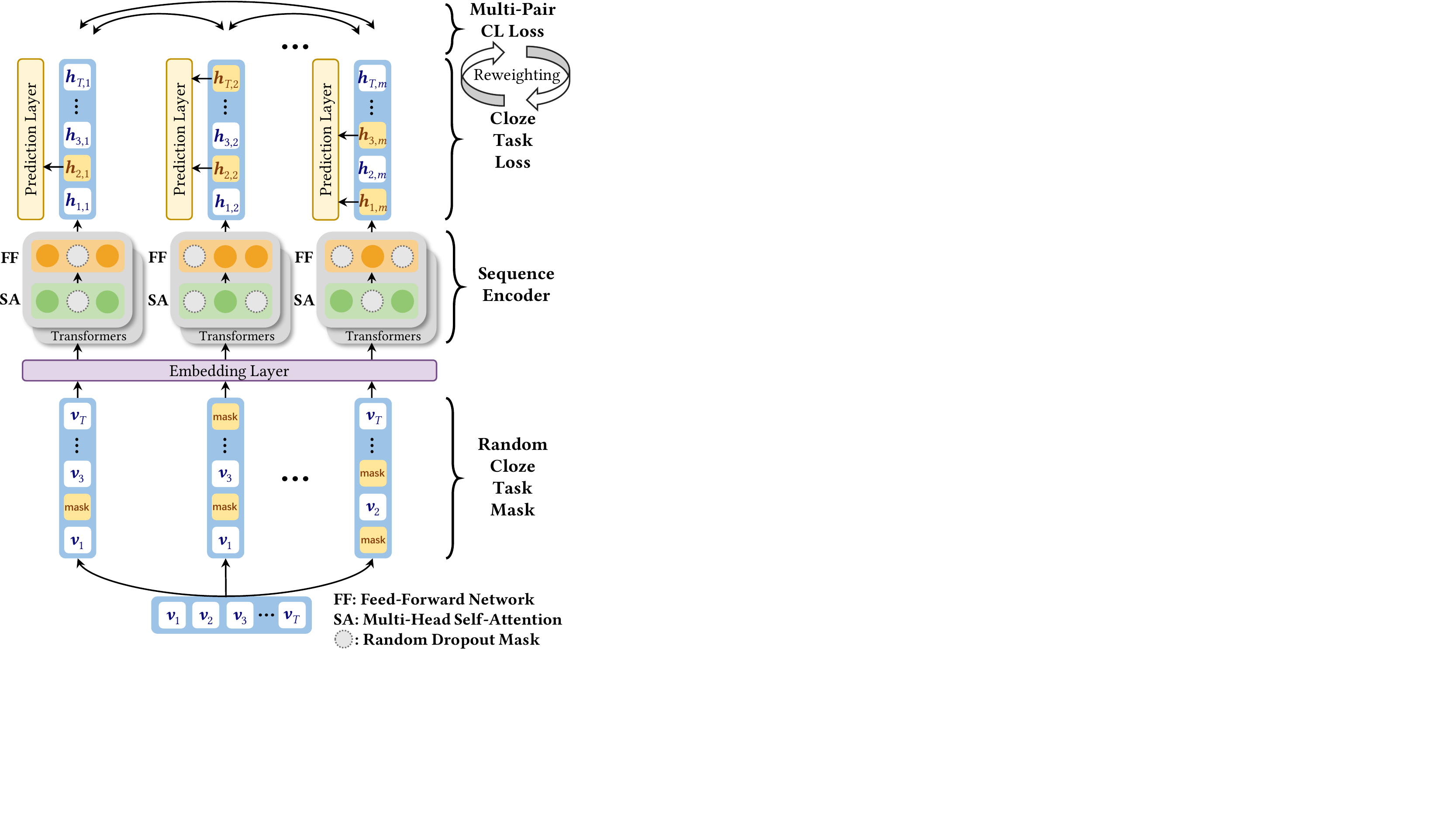}
\put(11.5,2.5){$s_u$}
\put(3.2,21){$s^{1}_{u}$}
\put(16.7,21){$s^{2}_{u}$}
\put(29.8,21){$s^{m}_{u}$}
\put(5.7,54.5){${\boldsymbol{H}}^{1}_{u}$}
\put(19.3,54.5){${\boldsymbol{H}}^{2}_{u}$}
\put(32.5,54.5){${\boldsymbol{H}}^{m}_{u}$}
\end{overpic}
 \caption{The architecture of CBiT. User sequence $s_u$ goes through random cloze task mask and generates $m$ different masked sequences. After passing through the embedding layer, each masked sequence is then forwarded through bidirectional Transformers with different random dropout masks. The masked items are used for the cloze task while the hidden representations of the whole sequence are used for contrastive learning. ${\boldsymbol{H}}^{1}_{u}, {\boldsymbol{H}}^{2}_{u}, \cdots, {\boldsymbol{H}}^{m}_{u}$ are all considered positive samples stemming from the same sequence $s_u$. Dynamic loss reweighting strategy is adopted to balance the cloze task loss and the multi-pair contrastive loss.}\label{modelfigure}\end{figure}
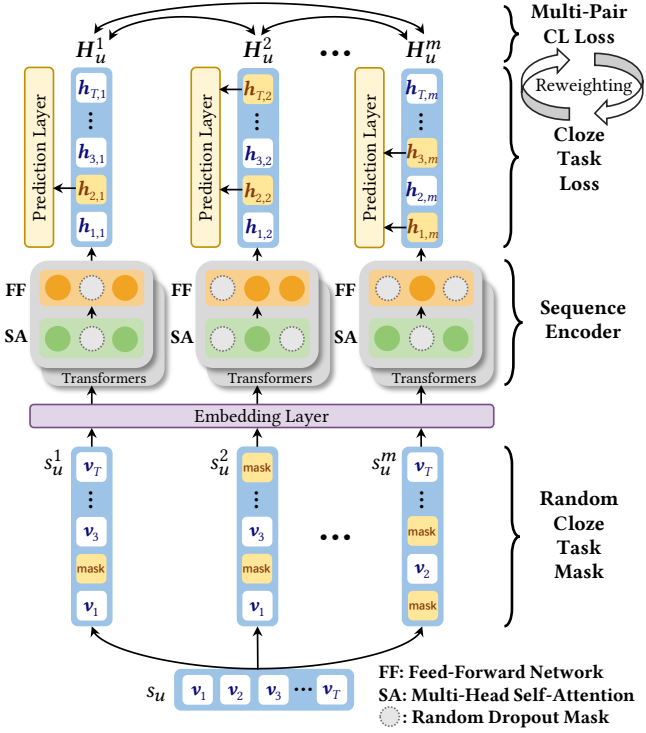
In this section, we introduce the architecture of our proposed framework, Contrastive learning with Bidirectional Transformers for sequential recommendation (CBiT) in detail. The architecture of CBiT is illustrated in Figure \ref{modelfigure}.  We first generate $m$ different masked sequences for each user sequence $s_u$. An embedding layer converts these masked sequences into embedding vectors. These embedding vectors are forwarded through bidirectional Transformers and the final outputs from the last layer are fetched as hidden representations. The cloze task is introduced as the main training objective, which requires the model to reconstruct the masked items based on the corresponding hidden representations of the masked items. The hidden representations of all the masked sequence are viewed as a collection of positive samples for multi-pair contrastive learning. Dynamic loss reweighting strategy is introduced to balance the main cloze task loss and the multi-pair contrastive loss.

\subsection{Base Model}
The base model of CBiT adopts an embedding layer and bidirectional Transformers as sequence encoder to generate hidden representations of sequences. A simple linear network is adopted as the prediction layer to convert hidden representations of sequences into the probability distribution of candidate items. 
\subsubsection{Embedding Layer.}
In CBiT, an item embedding matrix ${\emph{\textbf{E}}}\in {\mathbb{R}}^{\mathcal{V}\times d}$ and a positional embedding matrix ${\emph{\textbf{P}}}\in {\mathbb{R}}^{T\times d}$
are combined together to construct hidden representations of sequences. Here $T$\footnote{For any short sequence with length $\lvert{s_u}\rvert<T$, we add $T-\lvert{s_u}\rvert$ padding token(s) before it.} denotes the maximum sequence length of our model and $d$ denotes the hidden dimensionality. Therefore, given an item $v_i$, its input representation can be denoted as follows:
\begin{equation}
\emph{\textbf{h}}^{0}_{i}=\emph{\textbf{e}}_{i}+\emph{\textbf{p}}_{t}, \quad 1{\leq}t{\leq}T
\end{equation}
where $\emph{\textbf{e}}_{i}\in {\emph{\textbf{E}}}$, $\emph{\textbf{p}}_{t}\in {\emph{\textbf{P}}}$ are the embedding vectors corresponding to item $v_i$ and position $t$, respectively.

Different from previous works, we resolve the restriction imposed from the maximum sequence length $T$ by sliding a window of size $T$ over any long sequence $s_u$ with length $\lvert s_u \rvert>T$. With this slide window technique, we can preserve all the training data as well as divide sequences at a more fine-grained level.

\subsubsection{Sequence Encoder}
After passing through the embedding layer, we stack $\emph{\textbf{h}}^{0}_{i}$ together into matrix ${\emph{\textbf{H}}}^{\thinspace 0}\in {\mathbb{R}}^{T\times d}$ as the hidden representation of the whole sequence. Supposing we have ${\emph{\textbf{H}}}^{\thinspace 0}=[\emph{\textbf{h}}^{0}_{1},\emph{\textbf{h}}^{0}_{2},\cdots,\emph{\textbf{h}}^{0}_{T}]$, we forward ${\emph{\textbf{H}}}^{\thinspace 0}$ through $L$ layers of bidirectional Transformer blocks, and the procedure is defined as follows:
\begin{equation}
{\emph{\textbf{H}}}^{\thinspace l}=\mathtt{Trm}({\emph{\textbf{H}}}^{\thinspace l-1}),\quad\forall \thinspace l\in[1,\cdots,L].
\end{equation}
The output of the hidden representation ${\emph{\textbf{H}}}^{\thinspace L}=[\emph{\textbf{h}}^{L}_{1},\emph{\textbf{h}}^{L}_{2},\cdots,\emph{\textbf{h}}^{L}_{T}]$ from the last layer $L$ is fetched as the final output of bidirectional Transformer encoder. For simplicity we will omit the superscript $L$ and use  ${\emph{\textbf{H}}}=[\emph{\textbf{h}}_{1},\emph{\textbf{h}}_{2},\cdots,\emph{\textbf{h}}_{T}]$ to denote the final output of bidirectional Transformer encoder in the following passage.
\subsubsection{Prediction Layer}Given the final output of any hidden representation $\emph{\textbf{h}}_{t}$ at position $t$, we adopt a simple linear layer to convert $\emph{\textbf{h}}_{t}$ into probability distribution over candidate items:
\begin{equation}
\label{prob}
P(v)=\textbf{\emph{W}}^{P}\emph{\textbf{h}}_{t}+\emph{\textbf{b}}^{P},
\end{equation}
where $\textbf{\emph{W}}^{P}\in {\mathbb{R}}^{\mathcal{\lvert V \rvert}\times d}$ is the weight matrix and $\textbf{\emph{b}}^{P}\in \mathbb{R}^{\mathcal{\lvert V \rvert}}$ is the bias term for the prediction layer.

We adopt a simple linear layer rather than a feed-forward network with item embedding matrix \cite{Sun2019bert} because in practise we find out that a prediction layer with a shared item embedding matrix will interfere with the contrastive learning task, which heavily relies on the shared item embedding matrix to calculate item similarity. Apart from saving computational cost, a prediction layer without an item embedding matrix also decouples the dependency between the cloze task and the contrastive learning task.
\subsection{Learning with The Cloze Task}To train bidirectional Transformers, the cloze task is introduced. For each iteration step, given sequence $s_u$, we generate $m$ masked sequences $s^{1}_{u},s^{2}_{u},\cdots,s^{m}_{u}$ using different random seeds. In each masked sequence $s^{j}_{u}(1{\leq}j{\leq}m)$, a proportion $\rho$ of all items in the sequence $s_u$ are randomly replaced with the mask token [mask], and the position indices of the masked items are denoted as ${\mathcal{I}}^{j}_{u}$. The model is required to reconstruct the masked items. The loss function for the cloze task as the main training objective is defined as follows:
\begin{equation}
\mathcal{L}_{\rm{main}}=-\sum_{j=1}^{m}\sum_{t\in{{\mathcal{I}}^{j}_{u}}}\left[\log\sigma({P(v_t|s^{j}_{u})})+{\sum_{v^{-}_{t}\notin s_u}\log 1-\sigma({P(v^{-}_{t}|s^{j}_{u})})}\right],
\end{equation}
where $\sigma$ is the sigmoid function, and the probability $P(\cdot)$ is defined as Equation \ref{prob}. Each ground-truth item $v_t$ is paired with one negative item $v^{-}_{t}$ that is randomly sampled. Note that we only consider the masked items when calculating the loss function for the cloze task.
\label{cloze}
\subsection{Multi-Pair Contrastive Learning}
\subsubsection{The Simple One-Pair Instance.} Contrastive learning aims to bring positive samples close to each other while pushing negative samples apart from positive samples. Normally, given a batch of sequences ${\{s_u\}}^{N}_{u=1}$ with batch size $N$, a pair of hidden representations ${\boldsymbol{H}}^{x}_{u}$, ${\boldsymbol{H}}^{y}_{u}$ stemming from the same original sequence $s_u$ are brought together as a pair of positive samples\footnote{Here $x$ and $y$ denote the indices of two different masked sequences. $x$ and $y$ satisfy $1{\leq}x,y{\leq}m$.} while the other $2(N-1)$ hidden representations from the same batch are considered negative samples \cite{chen2020simple}. The contrastive learning loss for one pair based on InfoNCE \cite{InfoNCE} can be defined as follows:
\begin{equation}
    \label{onepair}
    \ell({\boldsymbol{H}}^{x}_{u},{\boldsymbol{H}}^{y}_{u})=-\log{\frac{\textbf{e}^{<{\boldsymbol{H}}^{x}_{u},{\boldsymbol{H}}^{y}_{u}>/\tau}}{\textbf{e}^{<{\boldsymbol{H}}^{x}_{u},{\boldsymbol{H}}^{y}_{u}>/\tau}+\sum^{N}_{k=1,k\neq u}\sum_{c\in\{x,y\}}\textbf{e}^{<{\boldsymbol{H}}^{x}_{u},{\boldsymbol{H}}^{c}_{k}>/\tau}}},
\end{equation}
where $\tau$ is a temperature hyper-parameter. The cosine similarity function $<\boldsymbol{\phi}_1,\boldsymbol{\phi}_2>={\boldsymbol{\phi}_1}^{\top}\cdot\boldsymbol{\phi}_2/||\boldsymbol{\phi}_1||{\cdot}||\boldsymbol{\phi}_2||$ is adopted to calculate the similarity between two hidden representations. In practice, we find out that the cosine similarity function performs better than the dot product similarity function.
\subsubsection{Extrapolation.} Simply using one pair of positive samples for contrastive learning does not suffice to fully exploit the great potential of bidirectional Transformers, because: \emph{\textbf{a}}\textbf{)} the difference between one pair of positive samples might be diminutive, so the model could hardly learn anything useful with these easy positive samples; \emph{\textbf{b}}\textbf{)} there might exist false negative samples within one batch, which could hurt performance if not neutralized with more positive samples. Therefore, we extrapolate the standard contrastive learning pair to multi-pair instances. 

Recall that in the cloze task we generate $m$ masked sequences using different random seeds. Given the final output of the $m$ hidden representations ${\boldsymbol{H}}^{1}_{u}, {\boldsymbol{H}}^{2}_{u}, \cdots, {\boldsymbol{H}}^{m}_{u}$ corresponding to masked sequences $s^{1}_{u},s^{2}_{u},\cdots,s^{m}_{u}$, we combinatorially brought these hidden representations together as positive samples. The multi-pair contrastive loss function for $m$ positive samples is defined as follows:
\begin{equation}
    \label{multipair}
    \mathcal{L}_{\rm{cl}}=\sum_{x=1}^{m}\sum_{y=1}^{m}{\vmathbb{1}}_{[x\neq y]}\ell({\boldsymbol{H}}^{x}_{u},{\boldsymbol{H}}^{y}_{u}),
\end{equation}
where ${\vmathbb{1}}_{[x\neq y]}\in \{0,1\}$ is an indicator function evaluating to $1$ iff $x\neq y$.

Multi-pair contrastive learning increases both the number of positive samples and the number of negative samples. The total number of positive samples is extended to $m$ because we consider the hidden representations of $m$ masked sequences as positive samples. The total number of negative samples is extended to $m(N-1)$ because each positive sample brings the other $(N-1)$ samples from the same batch as negative samples.

Multi-pair contrastive learning answers the aforementioned fundamental issues: (1) \emph{How to choose an augmentation strategy for contrastive learning?} To generate ${\boldsymbol{H}}^{1}_{u}, {\boldsymbol{H}}^{2}_{u}, \cdots, {\boldsymbol{H}}^{m}_{u}$ we need to forward different masked sequences $m$ times through bidirectional Transformers, and at each forward pass the random dropout mask is different. That means each hidden representation is yielded from both a unique cloze task mask and a unique dropout mask. Such augmentation strategy implicitly utilizes the cloze task mask from data level and the dropout mask from model level, maximizing the difference between these positive samples. (2) \emph{How to construct reasonable positive samples for contrastive learning?} In fact, multi-pair contrastive learning manages to construct reasonable positive samples through the additional self-supervision signal from the cloze task. ${\boldsymbol{H}}^{1}_{u}, {\boldsymbol{H}}^{2}_{u}, \cdots, {\boldsymbol{H}}^{m}_{u}$ all stem from the original sequence, and the objective of the cloze task is to reconstruct the masked items. Therefore, these positive samples should share semantic similarity.

\subsection{Training and Inference}
\subsubsection{Overall Training Objective.}
Generally speaking, contrastive learning methods for sequential recommendation jointly minimize the main loss $\mathcal{L}_{\rm{main}}$ and the contrastive learning loss $\mathcal{L}_{\rm{cl}}$, which can be denoted as follows:
\begin{equation}
    \mathcal{L}_{\rm{joint}}=\mathcal{L}_{\rm{main}}+\theta\mathcal{L}_{\rm{cl}},
\end{equation}
where $\theta$ is a weighting hyper-parameter.

In contrast with these methods, which employ a static proportional hyper-parameter to indicate the significance of the contrastive learning loss, we design a simple yet effective strategy $\theta(\alpha, \lambda)$  for contrastive learning loss reweighting as follows:
\begin{equation}
\begin{aligned}
    \theta_{\sigma+1}&=\alpha\hat{\theta}+(1-\alpha)\theta_{\sigma},\\
    \hat{\theta}&=\frac{\mathcal{L}_{\sigma+1}{}_{\rm{main}}}{\mathcal{L}_{\sigma+1}{}_{\rm{main}} +\lambda \mathcal{L}_{\sigma+1}{}_{\rm{cl}}},
\end{aligned}
\end{equation}
where $\alpha$ is a "learning rate" hyper-parameter for contrastive loss proportion $\theta$, $\lambda$ is a rescaling factor which should be tuned for different types of the contrastive loss function, and $\mathcal{L}_{\sigma+1}{}_{\rm{main}}$ and $\mathcal{L}_{\sigma+1}{}_{\rm{cl}}$ denote the main cloze task loss and the contrastive learning loss at the $\sigma+1$-th iteration step, respectively.
$\theta$ is set to $0$ at first and will be updated when the training process for each iteration step is finished.
We cut off the gradient of $\mathcal{L}_{\sigma+1}{}_{\rm{main}}$ and  $\mathcal{L}_{\sigma+1}{}_{\rm{cl}}$ when computing $\theta_{\sigma+1}$, so it will not interfere with the standard back propagation procedure. Therefore, the joint loss function for our model at the $\sigma+1$-th iteration step can be written as follows:
\begin{equation}
    \mathcal{L}_{\sigma+1}{}_{\rm{joint}}=\mathcal{L}_{\sigma+1}{}_{\rm{main}}+\theta_{\sigma+1}(\alpha,\lambda)\mathcal{L}_{\sigma+1}{}_{\rm{cl}},
\end{equation}
\subsubsection{Inference}
At the inference stage, we append the mask token [mask] to the end of the sequence so that the model will predict the next item of this sequence:
\begin{equation}
    s^{'}_{u}=[v^{(u)}_1,v^{(u)}_2,\cdots,v^{(u)}_{\lvert s_u \rvert},[\rm{mask}]].
\end{equation}
The slide window technique is not needed at the inference stage. For any long sequence we directly truncate it to the last $T$ items.

\subsection{Comparison}
\begin{table}[!t]
    \centering
    \footnotesize
    \caption{Comparison with other contrastive learning models.}
    \label{slidewindowmodelcompare}
  \setlength{\tabcolsep}{1pt}{
    \begin{tabular}{lcccc}
    \hline
         
         &CL4SRec&CoSeRec&DuoRec&CBiT\\\hline
         Sequence Encoder&Uni&Uni&Uni&Bi\\
         Handle Long Sequence&Truncation&Truncation&Truncation&Slide Window\\
         Augmentation&Data&Data&Model&Hybrid\\
         Representation for CL&${\boldsymbol{H}}_{u}$&${\boldsymbol{H}}_{u}$&$\boldsymbol{h^{'}}={\boldsymbol{H}}_{u}[-1]$&${\boldsymbol{H}}_{u}$\\
         Number of Positive Samples&2&2&4&Up to $+\infty$\\
         Loss Reweighting&$\times$&$\times$&$\times$&$\checkmark$\\
         \hline
    \end{tabular}
}
\end{table}
%In this section, we compare CBiT to other contrastive learning frameworks in sequential recommendation as well as another bidirectional sequence encoder---BERT4Rec \cite{Sun2019bert}.\subsubsection{Comparison with other contrastive learning frameworks.}We compare our model with other other contrastive learning frameworks, including CL4SRec \cite{Xu2020Contrastive}, CoSeRec \cite{liu2021contrastive} and DuoRec \cite{DuoRec}. Differences are summarized in Table \ref{slidewindowmodelcompare} and described as follows:
We compare our model with other other contrastive learning frameworks, including CL4SRec \cite{Xu2020Contrastive}, CoSeRec \cite{liu2021contrastive} and DuoRec \cite{DuoRec}. Differences are summarized in Table \ref{slidewindowmodelcompare} and described as follows:

\textbf{Sequence Encoder.} A major difference is that CBiT adopts the bidirectional Transformers as sequence encoder while other models adopt the unidirectional Transformers as sequence encoders. Compared with unidirectional Transformres, the attention mechanism of bidirectional Transformers can capture behavioral patterns at a more fine-grained level (section \ref{slidewindow}).

\textbf{Handling Long Sequence.} Another difference is that for long user sequences, CBiT adopts the slide window technique while other models use the standard truncation technique. The slide window technique is able to preserve all the data and also allows for a more fine-grained division of user sequences (Section \ref{slidewindowanalysis}).

\textbf{Augmentation.} To perform contrastive learning, we need to apply augmentations to the original sequence in order to generate samples. Augmentations can be classified into two broad categories, i.e., data augmentation and model augmentation. CL4SRec and CoSeRec apply perturbations such as mask, reorder and substitution to the original sequence, which can be viewed as augmentations from the data level. DuoRec applies perturbations such as unsupervised dropout and supervised sampling, which can be viewed as augmentations from the model level. In CBiT the cloze task mask can be viewed as an implicit form of data augmentation and the dropout mask is augmentation from the model level, so CBiT can be viewed as a hybrid of data augmentation and model augmentation.

\textbf{Representation for Contrastive Learning.} A more subtle difference lies in choices of the hidden representations for contrastive learning. Most models employ the hidden representation of the whole user sequence ${\boldsymbol{H}}_{u}$ while only DuoRec adopts the hidden representation of the last item $\boldsymbol{h^{'}}={\boldsymbol{H}}_{u}[-1]$ in the user sequence. The former option focuses more on aligning the semantics of the whole user sequence while the latter emphasizes more on the next-item prediction task. It is hard to say which option is better depending on different types of augmentations.

\textbf{Number of Positive Samples.} The normal contrastive learning method maximizes the agreement between a pair of positive samples, which is the case in CL4SRec and CoSeRec. DuoRec pursues two contrastive learning objectives from both supervised and unsupervised perspective, so it has two pairs of positive samples. CBiT extrapolates the normal contrastive learning to multi-pair instances, so in theory it can handle infinite positive samples.

\textbf{Loss Reweighting Strategy.} A novel dynamic loss reweighting strategy is introduced in CBiT to balance the main cloze task loss and the contrastive learning loss. In practice, we find that this strategy leads to smooth loss convergence and better model performance (Section \ref{lossreweightinganalysis}).
%%\subsubsection{Comparison with BERT4Rec.}BERT4Rec \cite{Sun2019bert} also adopts bidirectional Transformers as sequence encoder. Other than the additional contrastive learning module in CBiT, there are still a few differences between BERT4Rec and CBiT: \emph{\textbf{a}}\textbf{)} The slide window technique is introduced in CBiT to resolve the restriction of maximum sequence length. \emph{\textbf{b}}\textbf{)} The prediction layer in CBiT is a simplified linear network without an item embedding matrix.
\section{EXPERIMENT}\label{experiment}
In this section, we present the details of our experiments and answer the following research questions (\textbf{RQs}):
\begin{itemize}
[leftmargin =  8pt,topsep=1pt]
\item\textbf{RQ1}: How does CBiT perform comparing with other state-of-the-art methods? (Section \ref{overallperformance})
\item\textbf{RQ2}: What are the influence of different hyper-parameters in CBiT? (Section \ref{hyperparametersensitivity})
\item\textbf{RQ3}: What are the effectiveness of various novel techniques in CBiT? (Section \ref{ablationstudy})
\item\textbf{RQ4}: Why can bidirectional Transformers outperform unidirectional Transformers by a large margin? (Section \ref{slidewindow})
\end{itemize}
\subsection{Settings}
\subsubsection{Dataset.}
\begin{table}
    \centering
        \footnotesize
    \caption{\textbf{Dataset statistics after preprocessing}}
    \label{dataset}
  \setlength{\tabcolsep}{5pt}{
    \begin{tabular}{cccccc}
    \hline
         Datasets&\#users&\#items&\#actions&avg.length&sparsity\\\hline
         Beauty&22,363&12,101&198,502&8.9&99.93\%\\
         Toys&19,412&11,924&167,597&8.6&99.95\%\\
         ML-1M&6,040&3,953&1,000,209&163.5&95.21\%\\\hline
    \end{tabular}}
\end{table}
We conduct experiments on three public benchmark datasets. The Amazon dataset \cite{mcauley15image} contains users' reviews on products from different domains, which have relatively short sequence lengths. We select two domains \textbf{Beauty} and \textbf{Toys} as two different experimental datasets from the Amazon dataset. Another dataset MovieLens-1M (\textbf{ML-1M}) \cite{movielens} contains users' ratings on movies, which has very long sequences. All interactions are considered implicit feedbacks. We discard duplicated interactions and sort each user's interaction in the chronological order so as to construct user sequences. Following \cite{Sun2019bert,liu2021contrastive,DuoRec}, we omit users with less than 5 interactions and items related with less than 5 users. The \textit{leave-one-out} evaluation strategy is adopted, holding out the last item for test, the second-to-last for validation, and the rest for training. The processed dataset statistics are presented in Table \ref{dataset}.

\subsubsection{Metrics.}
For fair comparisons we rank the prediction on the whole item set \cite{krichene2020sampled}. We report the score of top-$K$ Hit Ratio (HR@$K$) and Normalized Discounted Cumulative Gain (NDCG@$K$).
\subsubsection{Baselines.}
The following baselines are used for comparisons:
\begin{itemize}
[leftmargin =  8pt,topsep=1pt]
    \item\textbf{GRU4Rec} \cite{srnn2016}. It introduces GRU with the ranking loss function in session-based recommendation.
    \item\textbf{Caser} \cite{tang2018caser}. It applies CNNs from both horizontal and vertical perspective for personalized sequential recommendation. The slide window technique is adopted according to the original paper.
    \item\textbf{SASRec} \cite{kang18attentive}. It uses unidirectional Transformers as sequence encoder, which serves as the base model for contrastive learning methods including CL4SRec, CoSeRec, and DuoRec.
    \item\textbf{CL4SRec} \cite{Xu2020Contrastive}. It is the first to apply contrastive learning to sequential recommendation. Data augmentation is adopted to generate positive samples.
    \item\textbf{CoSeRec} \cite{liu2021contrastive}. It further improves CL4SRec by introducing robust data augmentation methods.
    \item\textbf{DuoRec} \cite{DuoRec}. It is a very strong baseline which mitigates the representation degeneration problem in contrastive learning.
    \item\textbf{BERT4Rec} \cite{Sun2019bert}. It uses bidirectional Transformer as sequence encoder and adopts the cloze task to train the model.
    \item$\textbf{BERT4Rec}_S$. For comparison, we apply the slide window technique to BERT4Rec, which is roughly equivalent to removing the contrastive learning module of CBiT.
    \item\textbf{CoSeBERT}. We devise a variant of CoSeRec using BERT4Rec as the sequence encoder. Contrastive learning is the same with CoSeRec. The slide window technique is also adopted.
\end{itemize}
\subsubsection{Implementation.}
For GRU4Rec, Caser, SASRec, CoSeRec, and DuoRec, we use the codes provided by their authors. We implement CL4SRec, BERT4Rec, $\rm{BERT4Rec}_{\emph{S}}$ and CoSeBERT in PyTorch \cite{pytorch}. The number of Transformer blocks and the number of attention heads are tuned from $\{1,2,4\}$. The dropout ratio is tuned from $0.1$ to $0.9$. For CL4SRec, CoSeRec and CoSeBERT, we tune the ratios for different types of data augmentations from $0.1$ to $0.9$. We follow the instructions from the original papers to set other hyper-parameters.

We also implement our method in PyTorch \cite{pytorch}. Both the number of Transformer blocks $L$ and the number of attention heads $h$ are set as 2, and the hidden dimension as well as the batch size are set as 256. We set the mask proportion $\rho$ of the cloze task as 0.15, which is recommended by the author of BERT \cite{Devlin2019BERT}. We use the Adam \cite{Adam} optimizer with a learning rate of $0.001$, $\beta_{1}=0.9$, $\beta_{2}=0.999$, and the learning rate will decay exponentially \cite{loshchilov2017decoupled} after every 100 epochs. For other hyper-parameters, we tune dropout ratio from 0.1 to 0.9, rescaling factor $\lambda$ from 1 to 9, $\tau$ from 0.1 to 6, the number of positive samples $m$ from 2 to 8, the slide window size $T$ from 10 to 100, and $\alpha$ within $\{0.0001,0.0005,0.001,0.05,0.1\}$. We train our model for 250 epochs and select the checkpoint which displays the best NDCG@10 score on the validation set for test.
\begin{table*}[ht]
  \centering
  \small
     \caption{Overall performance of different methods on next-item prediction. The best score and the second best score in each row are bolded and underlined, respectively. Improvements over the best baseline method are indicated in the last column.}
  \renewcommand\arraystretch{1}  
  \setlength{\tabcolsep}{3.3pt}{

    \begin{tabular}{clccccccccccc}
\hline    
Dataset&Metric&GRU4Rec&Caser&SASRec &CL4SRec&CoSeRec&DuoRec&BERT4Rec&$\rm{BERT4Rec}_{\emph{S}}$&CoSeBERT&CBiT&Improv. \\
\hline   \multirow{6}[0]{*}{Beauty} 
&HR@5&0.0206&0.0254&0.0371&0.0396&0.0504&\underline{0.0559}&0.0370&0.0513&0.0546&\textbf{0.0637}& 13.95\%\\
&HR@10  &0.0332&0.0436&0.0592&0.0630&0.0726&\underline{0.0867}&0.0598&0.0755  &0.0786&\textbf{0.0905}& 4.38\%\\
&HR@20  &0.0526&0.0682&0.0893&0.0965&0.1035&\underline{0.1102}&0.0935&0.1068  &0.1093&\textbf{0.1223}& 10.98\%\\
&NDCG@5 &0.0139&0.0154&0.0233&0.0232&0.0339&0.0331&0.0233&0.0353  &\underline{0.0378}&\textbf{0.0451}& 19.31\%\\
&NDCG@10&0.0175&0.0212&0.0284&0.0307&0.0410&0.0430&0.0306&0.0431  &\underline{0.0462}&\textbf{0.0537}& 16.23\%\\
&NDCG@20&0.0221&0.0274&0.0361&0.0392&0.0488&0.0524&0.0391&0.0510  &\underline{0.0529}&\textbf{0.0617}& 16.64\%\\
\hline  \multirow{6}[0]{*}{Toys} 
&HR@5 &0.0121&0.0205& 0.0429  & 0.0503 & 0.0533 &0.0539&0.0371   & 0.0520 & \underline{0.0542} & \textbf{0.0640}& 18.08\%\\
&HR@10 &0.0184&0.0333& 0.0652 & 0.0736 & 0.0755 & 0.0744 &0.0524  &  \underline{0.0761} & 0.0735 & \textbf{0.0865}  & 13.67\%\\
&HR@20 &0.0290&0.0542& 0.0957  & 0.0990 & 0.1037 & 0.1008 & 0.0760 & \underline{0.1040} & 0.1019 & \textbf{0.1167}  & 12.21\%\\
&NDCG@5 &0.0077&0.0125& 0.0248 & 0.0264 & \underline{0.0370} &0.0340&0.0259 &0.0366 & 0.0355 &  \textbf{0.0462} & 24.86\%\\
&NDCG@10 &0.0097&0.0168& 0.0320 & 0.0339 & \underline{0.0442} &0.0406& 0.0309& 0.0442 & 0.0434 &\textbf{0.0535} & 21.04\%\\
&NDCG@20 &0.0123&0.0221& 0.0397 & 0.0404 & 0.0513 &0.0472& 0.0368& \underline{0.0521} & 0.0506 &\textbf{0.0610} & 17.08\%\\
\hline    
  \multirow{6}[0]{*}{ML-1M} 
&HR@5 &0.0806&0.0912&  0.1078  & 0.1142 & 0.1128 & \underline{0.1930} & 0.1308 & 0.1800 & 0.1653 & \textbf{0.2095}  &  8.55\%\\
&HR@10  &0.1344&0.1442& 0.1810 & 0.1815 & 0.1861 & \underline{0.2865} & 0.2219 & 0.2594 & 0.2492 & \textbf{0.3013} &  5.17\%\\
&HR@20  &0.2081&0.2228& 0.2745 & 0.2818 & 0.2950 & \underline{0.3901} & 0.3354 &  0.3623 & 0.3463 & \textbf{0.3998}  & 2.49\%\\
&NDCG@5  &0.0475&0.0565& 0.0681 & 0.0705 & 0.0692 & 
\underline{0.1327} & 0.0804 & 0.1215 & 0.1156 & \textbf{0.1436} & 8.21\%\\
&NDCG@10  &0.0649&0.0734& 0.0918 & 0.0920 & 0.0915 & \underline{0.1586}& 0.1097 & 0.1471 & 0.1412 & \textbf{0.1694}  & 6.81\%\\
&NDCG@20  &0.0834&0.0931& 0.1156 & 0.1170 & 0.1247 & \underline{0.1843} &0.1384 & 0.1729 & 0.1658 & \textbf{0.1957}  & 6.19\%\\
\hline\end{tabular}}
  \label{result}
\end{table*}
\subsection{Overall Performance Comparisons\label{overallperformance}}
Table \ref{result} presents the overall performance of our model and other baselines on the three public benchmark datasets. 

For comparisons of base sequence encoders, bidirectional Transformers with the slide window technique (e.g. $\rm{BERT4Rec}_{\emph{S}}$) outperforms models with other types of sequence encoders such as RNN (e.g. GRU4Rec), CNN (e.g. Caser) and unidirectional Transformers (e.g. SASRec) by a large margin. However, bidirectional Transformers with the normal truncation technique (e.g. BERT4Rec) performs similarly with its unidirectional counterpart (e.g. SASRec).

For comparisons of contrastive learning methods, models based on model augmentation (e.g. DuoRec) or hybrid augmentation (e.g. CBiT) outperform models based on data augmentation methods (e.g. CL4SRec, CoSeRec and CoSeBERT). We conjecture that data augmentation might corrupt the original semantics of sequences, and it also collides with the cloze task in bidirectional Transformers.

The performance improvement of our model can be attributed to two parts, i.e., the gain from the slide window technique and the gain from multi-pair contrastive learning. Our model also demonstrates a good adaptability and achieve the best performance on both short-sequence datasets and long-sequence datasets.
\subsection{Hyper-parameter Sensitivity\label{hyperparametersensitivity}}In this section, we study the influence of four important hyper-parameters in CBiT, including the number of positive samples $m$, the dropout ratio, temperature $\tau$, and the slide window size $T$. To control variables we only change one hyper-parameter at one time while keeping others optimal.
\begin{figure*}
\centering
\subcaptionbox{Number of positive samples\label{numpositive}}
{\includegraphics[width=.245\linewidth, trim=0 16 0 0, clip]{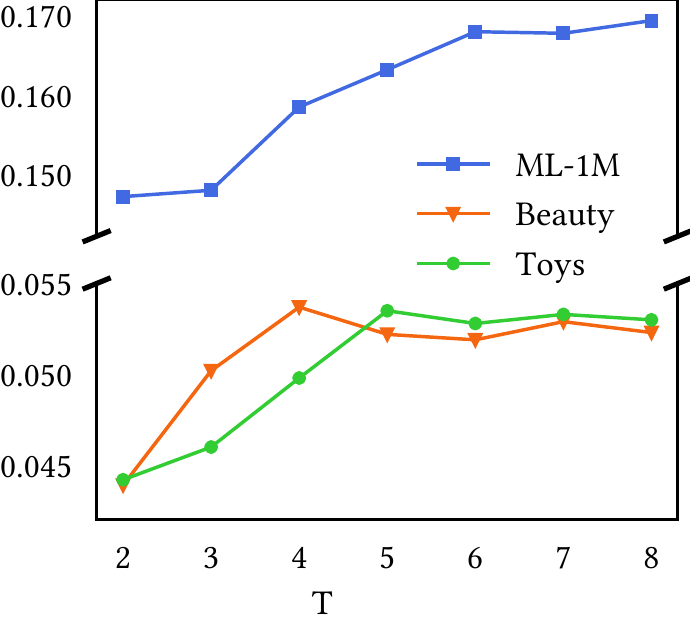}}
\subcaptionbox{Dropout ratio\label{dropout}}
{\includegraphics[width=.245\linewidth, trim=0 18 0 0, clip]{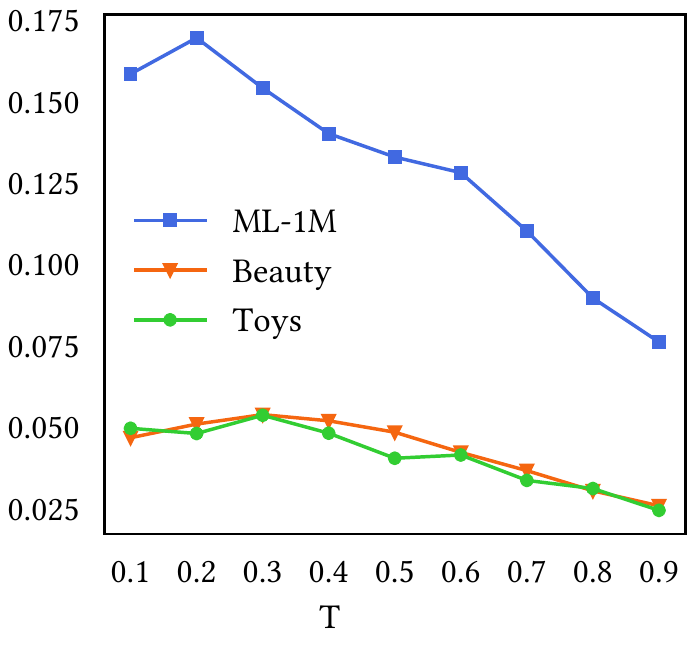}}
\subcaptionbox{Temperature\label{temperature}}
{\includegraphics[width=.245\linewidth, trim=0 16 0 0, clip]{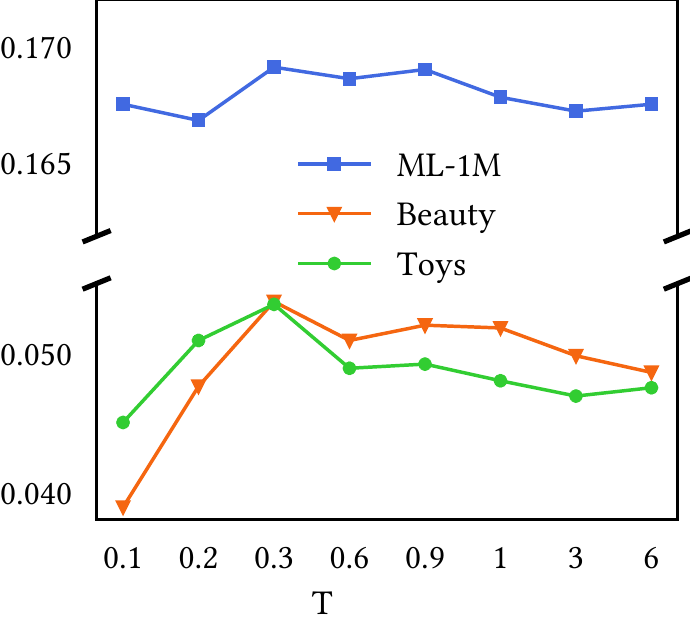}}
\subcaptionbox{Slide window size\label{seqlen}}
{\includegraphics[width=.245\linewidth, trim=0 16 0 0, clip]{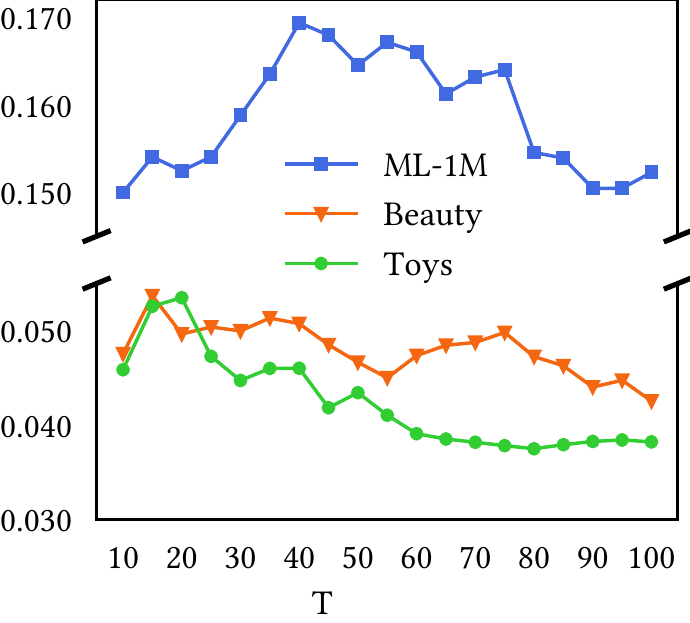}}
\caption{Performance (NDCG@10) comparison w.r.t different hyper-parameters on three datasets.}\label{hyperparameter}
\end{figure*}
\subsubsection{Number of Positive Samples.} The number of positive samples $m$ regulates how many positive samples are available for multiple-pair contrastive learning. Figure \ref{numpositive} shows that increasing the number of positive samples can improve performance. Contrastive learning with multiple positive samples performs better than the simple contrastive learning with a pair of positive samples. This phenomenon can be attributed to the variety of semantic patterns brought about by multiple positive samples. However, the effectiveness of multiple positive samples also has an upper limit. When there are sufficient positive samples, performance reaches a plateau even if we add more positive samples.\label{numberofpositivesamples}
\subsubsection{Dropout Ratio.} On the one hand, the dropout ratio should not be too large so as to avoid corrupting the original semantics of sequences. On the other hand, we hope to adopt a larger dropout ratio so as to generate harder samples for the contrastive learning task. Therefore, we have to strike a balance when choosing the optimal dropout ratio. As we can see in Figure \ref{dropout}, CBiT is unable to reach its best performance when the dropout ratio is 0.1. A large dropout ratio such as 0.9 also significantly cuts down performance. The optimal dropout ratio should be somewhere in between, which is 0.3 for Beauty and Toys, 0.2 for ML-1M in CBiT.
\subsubsection{Temperature.} Temperature $\tau$ regulates how much we should penalize for hard negative samples \cite{wang2021understanding}. A small $\tau$ leads to an excessive pursuit of uniformity \cite{wang2020alignmentuniformity}, which may break underlying semantic similarity, while a large $\tau$ makes the contrastive loss function too soft to discriminate between samples. From Figure \ref{temperature} we can see that
a good choice of temperature should neither be too large nor too small, which is 0.3 in our model.
\subsubsection{Slide Window Size.\label{analysisonseqlen}} It is very important to choose an appropriate slide window size because the slide window size depends the maximum sequence length of our model. Comparing with the average length of training dataset, an appropriate slide window size $T$ for the model should not be too large, because: \emph{\textbf{a}}\textbf{)} only recent interactions play a significant role in indicating user's intent while outdated interactions are less helpful; \emph{\textbf{b}}\textbf{)} a large value $T$ results in too much padding, which will inevitably hurt performance.

Besides, using a smaller slide window size can also improve efficiency. The time complexity of CBiT for each layer is $\mathcal{O}(T^2d)$. By reducing $T$ to the near average sequence length of each dataset, we can greatly save computational cost.

From Figure \ref{seqlen} we can see that the optimal slide window size is 15 for Beauty, 20 for Toys, and 40 for ML-1M respectively. Datasets with shorter sequences favour a shorter slide window size whereas datasets with longer sequences favour a slightly longer sequence length. And even on ML-1M the optimal slide window size is not 160 but 40 because only recent interactions are more helpful when predicting the near future.

It has to be pointed out that using a shorter slide window size does not mean that outdated interactions should be considered as noise. Although they may not be helpful in predicting the future if users' current interests have shifted, we still need them in the training data because they give us insights into how users' behavioral patterns change dynamically. For instance, user A may be interested in items from group A in the past, but now his or her interest has shifted to items from group B. When predicting the future for user A, considering interactions from group A may not be helpful because of interest shift. However, if another user B happens to show interest in items from group A, then we can use information learned from user A to predict the future for user B.

\subsection{Ablation Study\label{ablationstudy}}In this section, we perform ablation study on the slide window technique, the augmentation strategy and the loss reweighting strategy of CBiT to understand their effectiveness.
\subsubsection{Slide Window Technique.}\label{slidewindowanalysis}
\begin{table}
    \centering
    \footnotesize
    \caption{Analysis on the effectiveness of the slide window technique (denoted as ${\emph{S}}$). Note that Caser already adopts the slide window technique according to the original paper.}
    \label{slidewindowmodel}
      \renewcommand\arraystretch{1}  
  \setlength{\tabcolsep}{2.8pt}{
    \begin{tabular}{lccccccc}
    \hline
         \multirow{2}[0]{*}{Model}&\multicolumn{2}{c}{Beauty}&\multicolumn{2}{c}{Toys}&\multicolumn{2}{c}{ML-1M}\\\cmidrule(lr){2-3}\cmidrule(lr){4-5}\cmidrule(lr){6-7}
         &HR@10&NDCG@10&HR@10&NDCG@10&HR@10&NDCG@10\\\cmidrule{1-7}
         GRU4Rec&0.0332&0.0175&0.0184&0.0097&0.1344&0.0649\\
         $\rm{GRU4Rec}_{\emph{S}}$&0.0237&0.0113&0.0101&0.0052&0.0990&0.0438\\\cmidrule{1-7}
         Caser&0.0436&0.0212&0.0333&0.0168&0.1442&0.0734\\\cmidrule{1-7}
         SASRec&0.0592&0.0284&0.0652&0.0320&0.1810&0.0918\\
         $\rm{SASRec}_{\emph{S}}$&0.0585&0.0278&0.0694&0.0339&0.1727&0.0845\\\cmidrule{1-7}
         BERT4Rec&0.0598&0.0306&0.0524&0.0309&0.2219&0.1097\\
         $\rm{BERT4Rec}_{\emph{S}}$&0.0755&0.0431&0.0761&0.0442&0.2594&0.1471\\\hline
    \end{tabular}
}
\end{table} To study its effectiveness, we use the slide window technique on different types of base sequence encoders and examine their performance. We tune the slide window size for each model and report the best performance in Table \ref{slidewindowmodel}. We can see that no other types of sequence encoders achieves such a remarkable performance improvement as bidirectional Transformers with the introduction of the slide window technique.

We attribute this phenomenon to the agility of the bidirectional attention mechanism in bidirectional Transformers. The attention mechanism in bidirectional Transformers can attend on items from both sides, which is beneficial for capture more personalized, fine-grained and nuanced behavioral patterns. Therefore, we can reduce the slide window size to a relatively small value, which means we divide a long sequence into a few sub-sequences where each sub-sequence displays slightly different behavioral patterns.
By feeding these sub-sequences into our model, bidirectional Transformers can capture more fine-grained features. This phenomenon can also be affirmed from the observation in Figure \ref{seqlen} that using a large slide window size undermines performance, which means that bidirectional Transformers have not yet fully capture more fine-grained behavioral patterns when long sequences have not been broken into sub-sequences.

By comparison, other sequence encoders cannot capture behavioral patterns at a more fine-grained level. So it does not make much difference even if we break sequences into sub-sequences.

\begin{figure}
\centering
\subcaptionbox{Augmentation strategy\label{positivesamples}}
{\includegraphics[ width=.49\linewidth, trim= 14 0 10.3 0,clip ]{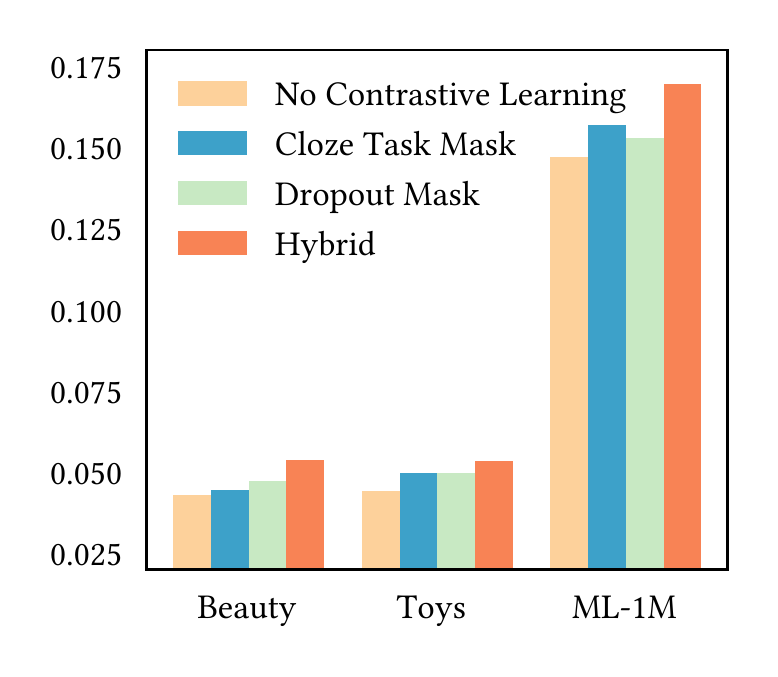}}
\subcaptionbox{Loss reweighting strategy\label{lossreweighting}}
{\includegraphics[ width=.49\linewidth ]{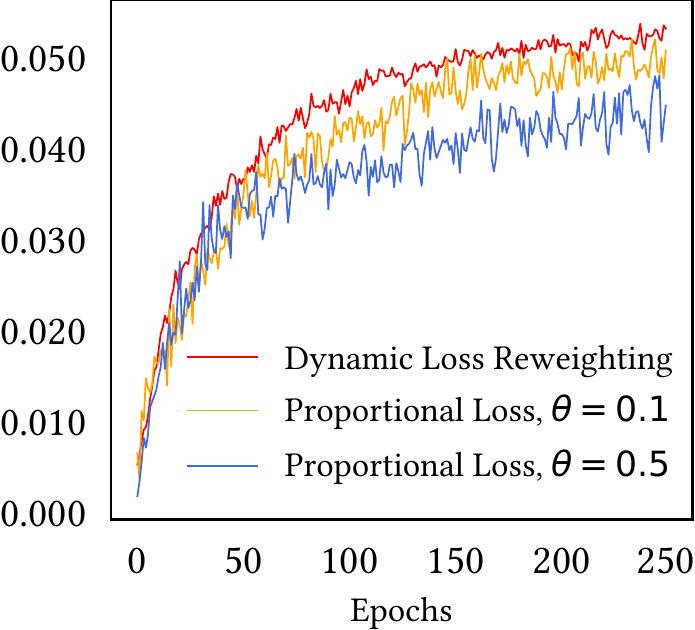}}
\caption{Ablation study (NDCG@10) w.r.t to the augmentation strategy and the loss reweighting strategy.}
\end{figure}
\subsubsection{Augmentation Strategy.} Our model uses both the cloze task mask and the dropout mask as a hybrid augmentation strategy. To study the effectiveness of these two augmentations separately, we conduct ablation experiments on our augmentation strategy, where we only modify our augmentation strategy while keeping other hyperparameters optimal. From Figure \ref{positivesamples} we can see that enabling contrastive learning is better than no contrastive learning at all, and neither using the cloze task mask nor the dropout mask alone can achieve performances comparable with our hybrid strategy.

We explain the good performance of the hybrid approach from two perspectives: \emph{\textbf{a}}\textbf{)} a hybrid of these two mask mechanisms brings more perturbation to the original sequence, so the samples for contrastive learning can be harder and of better quality; \emph{\textbf{b}}\textbf{)} compared with other data augmentation methods, these two mask mechanisms inherent in the original BERT architecture are more compatible with bidirectional Transformers, so the perturbation from these two mask mechanisms will not break semantic similarity.
\subsubsection{Loss Reweighting Strategy.} In our experiments, we observe that the contrastive learning loss may be very large at the inception of the training phase. It takes more time for the model to converge with a large contrastive learning loss. However, simply setting the weighting hyper-parameter $\theta$ to a small value cannot solve the problem, because in the latter stage the contrastive learning loss may become too small for the model to focus on contrastive learning. Therefore, we design the dynamic loss reweighting strategy, which automatically reweights the contrastive learning loss. 

The ablation study of the loss reweighting strategy is presented in Figure \ref{lossreweighting}, where we use different weighting strategies and save the checkpoints from every epoch during training and test them on the Beauty dataset. We can see that using dynamic loss reweighting strategy in CBiT can improve performance and displays a smoother training curve. Setting neither the proportion $\theta$ of contrastive learning loss to a small value 0.1 nor to a large value 0.5 can achieve comparable effectiveness.
\label{lossreweightinganalysis}

\subsection{Discussion: Why Bidirectional Transformers?}
\begin{figure}
\centering
\subcaptionbox{SASRec, layer1\label{1a}}
{\includegraphics[width=.49\linewidth]{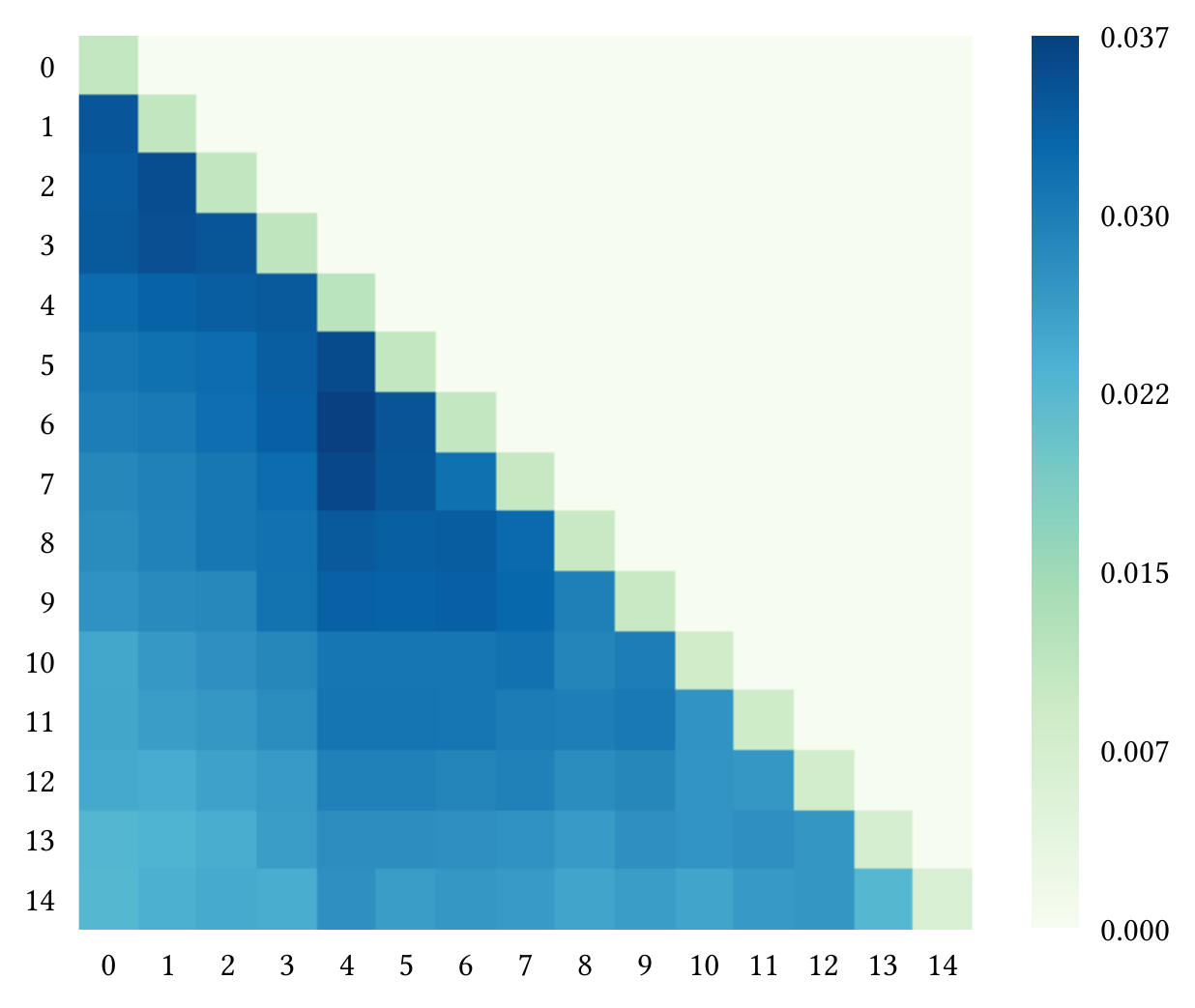}}
\subcaptionbox{SASRec, layer2\label{1b}}
{\includegraphics[width=.49\linewidth]{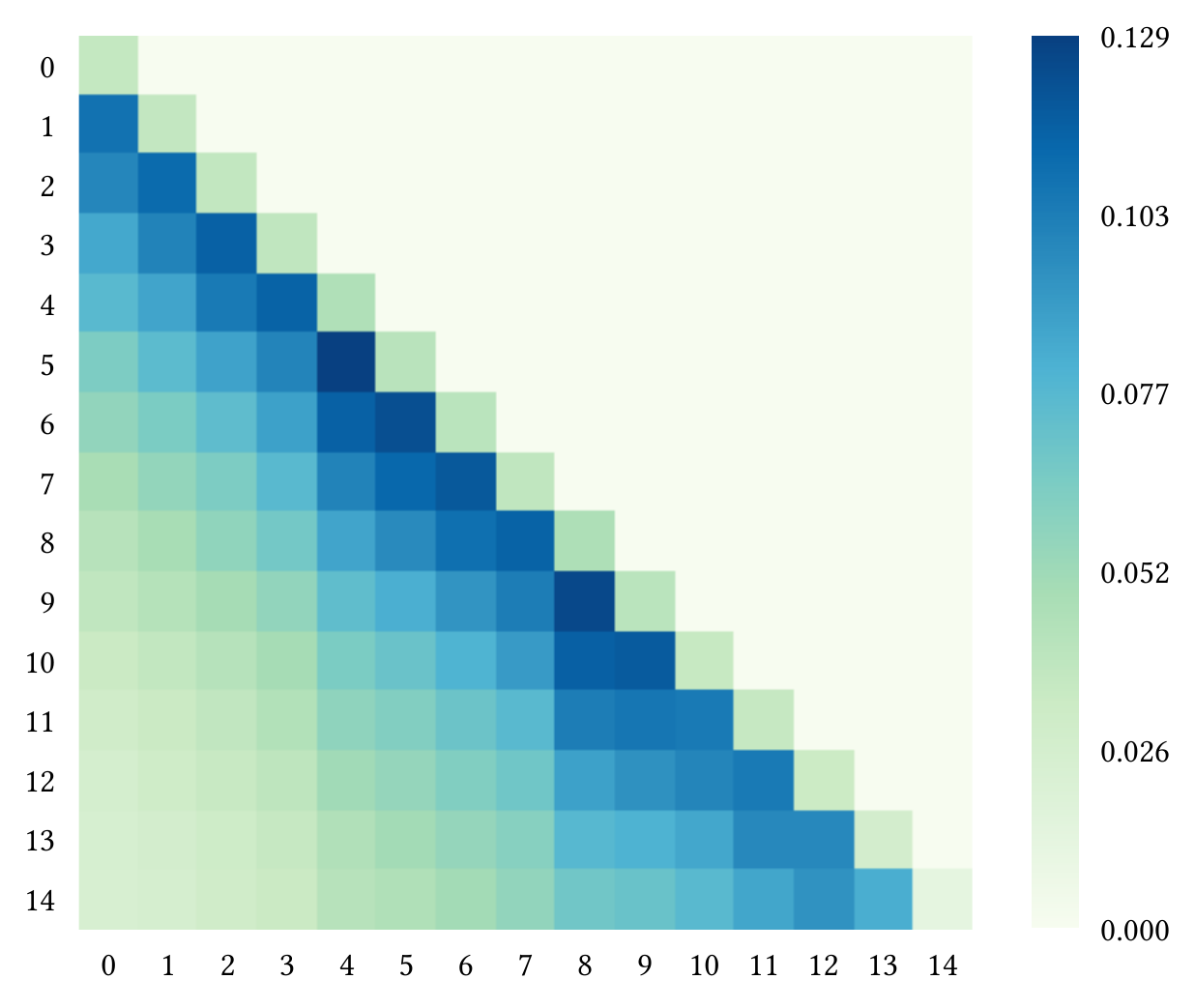}}
\subcaptionbox{CBiT, layer1\label{1c}}
{\includegraphics[width=.49\linewidth]{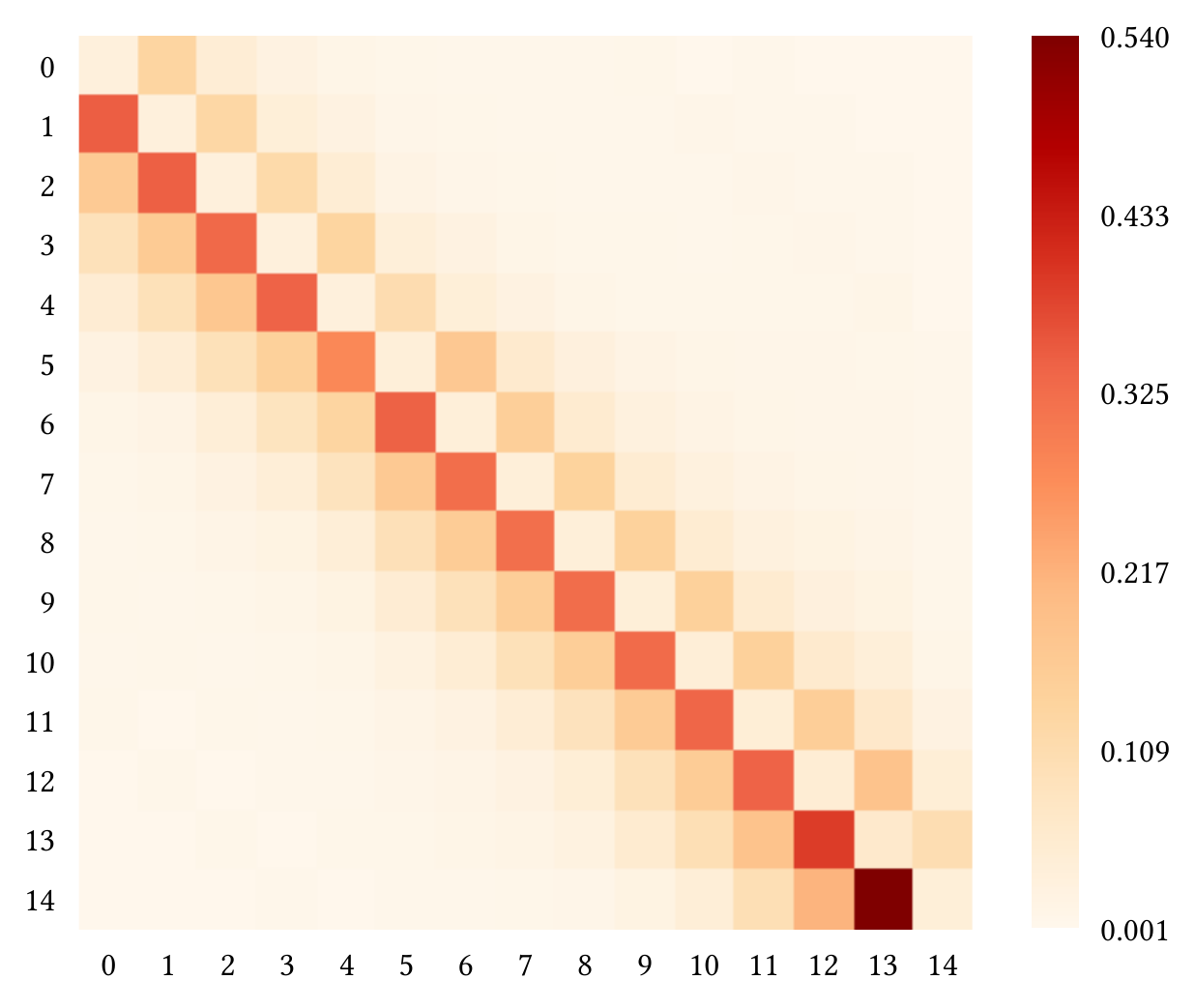}}
\subcaptionbox{CBiT, layer2\label{1d}}
{\includegraphics[width=.49\linewidth]{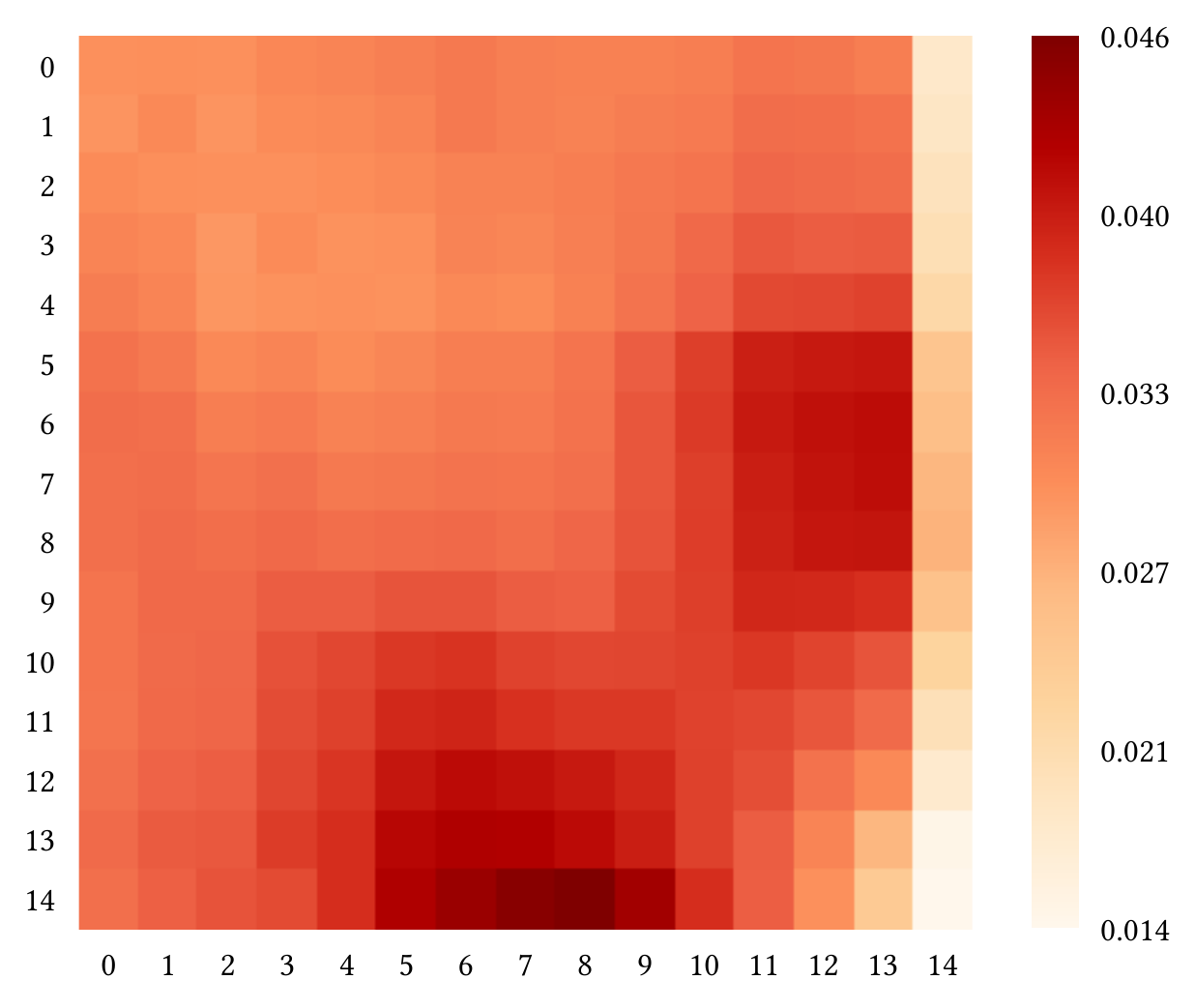}}
\caption{Average attention weights of SASRec and CBiT on the ML-1M dataset at different layers. In CBiT the last position represents the mask token. For simplicity, we only plot the average attention weights of the 2nd heads.}\label{attentionweights}\end{figure}

The main difference between unidirectional Transformers and bidirectional Transformers is whether self-attention can see future items in the current sequence. Unidirectional Transformers use the attention mask to hide future items, which creates a shifted version of the same sequence \cite{kang18attentive}. By comparisons, bidirectional Transformers have access to all the items within sequences, so it can see items from both sides \cite{Sun2019bert}. As we can see in Figure \ref{attentionweights}, for unidirectional Transformers the heat-map of average attention weights is diagonal because of such causality attention masking. This restricts its ability to consider subsequent context information because it can only see items from the left side. However, bidirectional Transformers can see items from both sides, which is beneficial for capturing more fine-grained behaviorial patterns.

The difference in the attention mechanisms leads to different training objectives. For unidirectional Transformers, the training objective is next-item prediction because the model learns to process sequence information from left to right. For bidirectional Transformers, the training objective is the cloze task using an additional mask token because the model learns to process sequence information from both sides. As we can see in Figure \ref{1c} and \ref{1d}, attention in the $2$-th head tends to attend on the mask token, which indicates that the model is sensitive to the cloze task.

Theoretically, bidirectional Transformers should perform better than unidirectional Transformers because of the difference in the attention mechanism. However, in our experiments we find out that bidirectional Transformers can only outperform unidirectional Transformers by a large margin on condition that the slide window technique is adopted, which indicates the necessity of using the slide window technique in bidirectional Transformers.
\label{slidewindow}
\section{CONCLUSION} In this work, we proposed a novel framework called Contrastive learning with Bidirectional Transformers for sequential recommendation (CBiT). We utilized both the cloze task mask and the dropout mask to generate multiple positive samples and extrapolate the normal one-pair contrastive learning to multi-pair instances. To smooth multi-pair contrastive loss, we designed a novel dynamic loss reweighting strategy. The slide window technique is also adopted to divide sequences from a more fine-grained level. Experimental results on the three public benchmark datasets showed that our approach outperforms several state-of-the-art methods and provides an insight into how and why our approach works.
\begin{acks}
This research was partially supported by the NSFC (61876117, 61876217, 62176175), the major project of natural science research in Universities of Jiangsu Province (21KJA520004), Project Funded by the Priority Academic Program Development of Jiangsu Higher Education Institutions and 
the exploratory self-selected subject of the SKLSDE.
\end{acks}
%%\section{Appendices}



\bibliographystyle{ACM-Reference-Format}
\balance
\bibliography{sample-base}

%%\appendix

\end{document}